\documentclass[rsi,reprint]{revtex4-1}
\usepackage{graphicx}
\usepackage{bm,braket,color,ulem}
\usepackage{amssymb,amsmath}

\begin{document}
\title{Force-detected high-frequency electron spin resonance spectroscopy using magnet-mounted nanomembrane: robust detection of thermal magnetization modulation}

\author{Hideyuki Takahashi$^{1}$\thanks{E-mail address: hide.takahashi@crystal.kobe-u.ac.jp},  Tsubasa Okamoto$^2$, Kento Ishimura$^2$, Shigeo Hara$^3$, Eiji Ohmichi$^{2}$, and Hitoshi Ohta$^{4}$}
\affiliation{
$^{1}$Organization for Advanced and Integrated Research, Kobe University, 1-1, Rokkodai, Nada, Kobe 657-8501, Japan\\
$^{2}$Graduate School of Science, Kobe University, 1-1 Rokkodai-cho, Nada, Kobe 657-8501, Japan\\
$^{3}$Research Facility Center for Science and Technology, Kobe University, 1-1 Rokkodai-cho Nada, Kobe 657-8501, Japan\\
$^{4}$Molecular Photoscience Research Center, Kobe University, 1-1 Rokkodai-cho, Nada, Kobe 657-8501, Japan}

\date{\today}

\begin{abstract}
In this study, we report a conceptually novel broadband high-frequency electron spin resonance (HFESR) spectroscopic technique.
In contrast to the ordinary force-detected ESR technique, which detects the magnetization change due to the saturation effect, this method measures the magnetization change due to the change of the sample temperature at resonance. 
To demonstrate its principle, we developed a silicon nitride nanomembrane-based force-detected ESR spectrometer, which can be stably operated even at high magnetic fields. 
Test measurements were performed for samples with different spin relaxation times. 
We succeeded in obtaining a seamless ESR spectrum in magnetic fields of 15~T and frequencies of 636~GHz without significant spectral distortion.
A high spin sensitivity of $10^{12}$~spins/G$\cdot$s was obtained, which was independent of the spin relaxation time.
These results show that this technique can be used as a practical method in research fields where the HFESR technique is applicable.
\end{abstract}

\maketitle
\section{Introduction}

Electron spin resonance (ESR) spectroscopy is a widely used technique in material science to study the local structure, electronic states, and spin dynamics. 
Conventional ESR spectrometers are operated below 100 GHz and commercially available. Their sensitivities ($\sim 10^9$ spins/G) are maintained by the use of the cavity resonators and established device technology~\cite{Poole}.
Nevertheless, there is a substantial need for measurements in the higher frequency region.
The so-called high-frequency ESR (HFESR) technique enables the study of several phenomena, which cannot be investigated by commercial spectrometers.
The examples include quantum spin materials~\cite{Haldane1983,Balents2010,Hikihara2008}, field-induced phase transitions~\cite{Ruette2004,Ohta1993}, and metal proteins with a large zero-field splitting~\cite{Okamoto2018}.

The spin sensitivity can be enhanced up to $\sim 10^{9}-10^{10}$ spins/G by using a Fabry--Perot resonator in the frequency range of 100--400 GHz~\cite{Earle1996,Tol2005,Morley2008}.
Among these, certain spectrometers are compatible with pulsed ESR experiments.
The only disadvantege of this method is that Fabry--Perot cavities are operated only around specific frequencies.
In HFESR experiments, one of the priorities is to obtain the spectra across wide frequency ranges and magnetic fields.
For this purpose, transmission-type measurements in pulsed or static magnetic fields have been preferred~\cite{Motokawa1991,Hassan2000}. 
The highest sensitivity achieved by the broadband method is $\sim 10^{10}-10^{12}$ spins/G, which was obtained by the combination of field modulation with a cryo-cooled InSb borometer~\cite{Hassan2000}.
However, it should be noted that the spin sensitivity depends on several variables, such as sample properties, the form of the sample (polycrystals or single crystals), sample dimensions, and experimental parameters (e.g. temperature, resonance frequency, microwave power, etc.)~\cite{Blank2017}.
The sensitivity value above is obtained under ideal conditions.
In order to successfully perform measurements regardless of these parameters, improving the sensitivity is important as well.

We believe that the force-detected ESR (FDESR) method has a potential to become a novel technique in the high-frequency region.
This method measures the change in magnetization $\bm{M}$ accompanying ESR by using a microcantilever~\cite{Rugar1992}. 
Its extraordinarily high sensitivity has been demonstrated in the radio frequency and microwave frequency regions~\cite{Kuehn2008}.
Although recently spatially resolved experiments are widely used, several attempts have demonstrated the suitability of the broadband feature to be applied in HFESR spectroscopy~\cite{Cruickshank2007,Ohmichi2008,Hallak2010,HT2015,Ohmichi2016}.
However, the force-detected technique has certain issues to be solved in order to establish it as a practical HFESR method.
One such problem is the difficulty of the application of the sensitive microcantilever. 
A microcantilever with a magnetic tip can deflect largely in a strong magnetic field, resulting in complex dynamics~\cite{Cruickshank2007,HT2016}.
The large deflection also makes the alignment with the laser beam difficult for displacement measurements. 
For these reasons, there have been no reports on obtaining seamless force-detected ESR spectrum successfully from 0~T to fields above 10~T.

A further problem is the limited range of applicable substances. In order to induce sufficient magnetization change, a number of spin-inverted electrons needs to be accumulated, which is only possible with substances with a long spin relaxation time. 
Therefore, the conventional FDESR technique has not been effective for magnetic materials with a very broad absorption linewidth.

In this study, we propose a sufficiently robust FDESR technique, which overcomes these issues.
The proposed technique indirectly measures the increase in temperature by the thermal dissipation in the process of spin relaxation, which is a concept different from those used in previous FDESR studies.
In addition, we report the novel type of FDESR spectrometer utilizing a silicon nitride (SiN$_x$) nanomembrane, which can achieve excellent stability and sensitivity even in strong magnetic fields above 10~T.
We demonstrate that the combination of these techniques is effective even for fast-relaxing spin systems.

\section{Principles}

\begin{figure}[tb]
	\begin{center}
		\includegraphics[width=0.95\hsize]{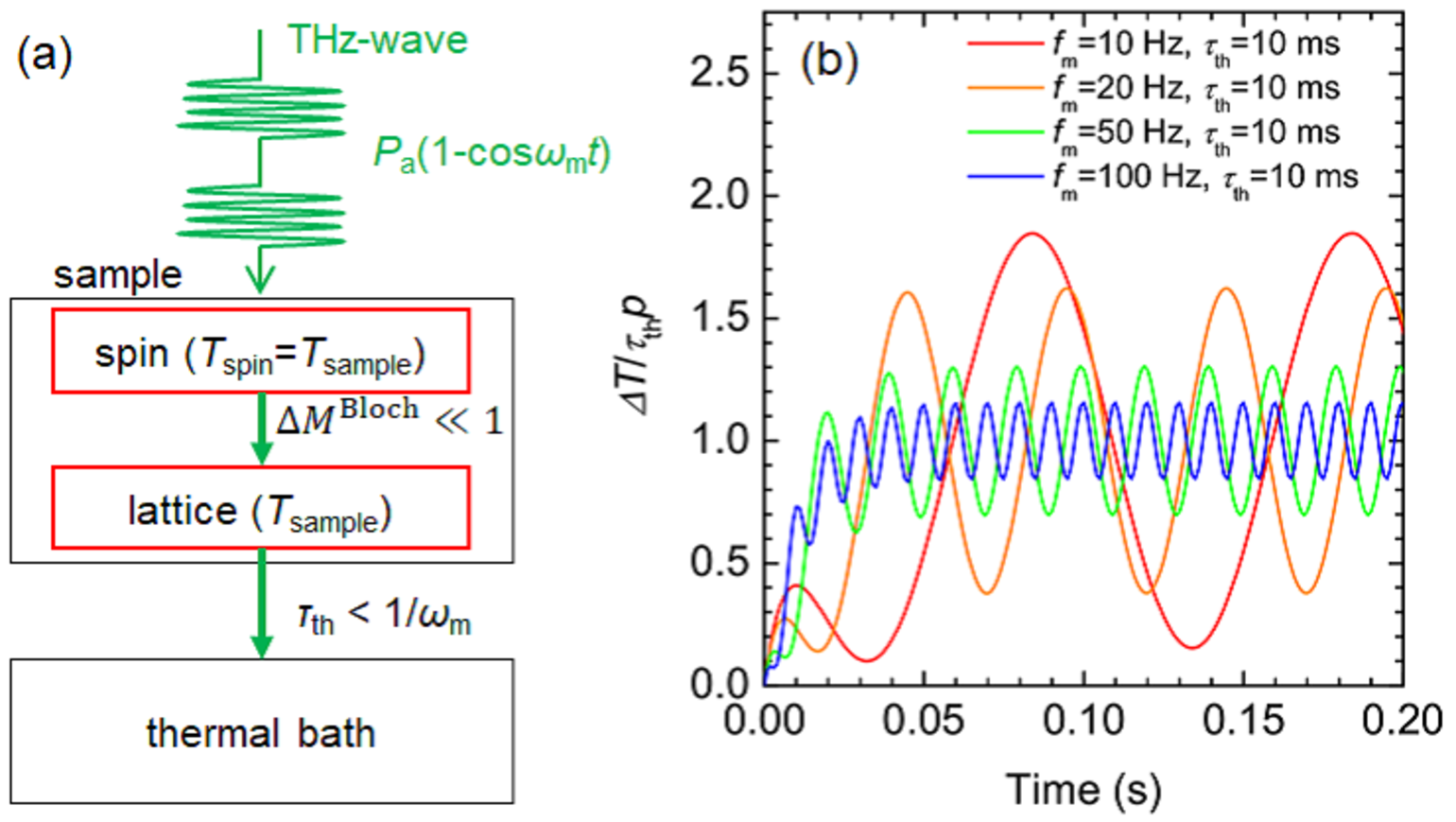}
	\end{center}
\caption{(a) Energy flow before and after ESR absorption for fast relaxing spin systems and (b) 
temperature modulation for $\tau_{th}=10\ \mathrm{ms}$ after the start of electromagnetic wave irradiation at $t=0$.}
\label{model}
\end{figure}

The FDESR signal is detected as the change in the longitudinal component of $\bm{M}$, $M_z$.
In a typical case, the magnetization change is attributed to the electron transition between different spin states, which is obtained by solving the Bloch equation~\cite{Abragam}:
\begin{equation}
\label{Bloch}
\Delta M_z^{\mathrm{Bloch}}=-M_0\frac{\gamma^2 B^2_1 T_1 T_2}{1+\gamma^2 B^2_1 T_1 T_2},
\end{equation}
where $M_0$ is the equilibrium magnetization, $B_1$ is the transverse magnetic field of the electromagnetic wave, and $\gamma$ is the gyromagnetic ratio. 
In the high-frequency region, $|\Delta M_z^{\mathrm{Bloch}}/M_{0}| \ll 1$ due to the weak power of the light source.
Even for 2,2-diphenyl-1-picrylhydrazyl (DPPH) with a relatively long $T_1=70\ \mathrm{ns}$ and $T_2=200\ \mathrm{ns}$, $|\Delta M_z^{\mathrm{Bloch}}/M_{0}|$ is estimated to be $4\times 10^{-4}$ with $B_1=10^{-6}\ \mathrm{T}$~\cite{DPPHT12}
In our setup shown in Sec. III, this $B_1$ value corresponds to the light power of 3 mW when the transmitted THz-wave is approximated by a plane wave.
In magnetic materials, $T_{1(2)}$ is often shorter than 1 ns due to the strong exchange interaction, hence $|\Delta M_z^{\mathrm{Bloch}}/M_{0}|$ is smaller by several orders.
This has been a fundamental obstacle in the application of the force-detection method to HFESR spectroscopy.

In this study, we focus on another type of magnetization modulation.
This mechanism is schematically shown in Fig~\ref{model}(a).
Here, we consider a low power measurement of fast-relaxing samples in which the temperature of the spin system can be assumed to be equal to the sample temperature regardless of its $T_{1(2)}$ values.
During ESR, the sample partially absorbs the power of the electromagnetic wave, $P_{a}(1-\cos\omega_m t)/2$, where $P_a$ is the peak value of the absorbed power and $\omega_m=2\pi f_m$ the amplitude modulation frequency.
Then, the absorbed energy immediately flows to the lattice system. 
After spin relaxation, the thermal energy stored in the sample is released to the thermal bath.
The time constant of this process $\tau_{th}$ is determined by the thermal conductivity of the sample and the thermal contact between the sample and the thermal bath.
The increase in $T_\mathrm{sample}$, $\Delta T$, satisfies the following differential equation,
\begin{equation}
\frac{d}{dt}\Delta T=-\frac{1}{\tau_{th}}\Delta T+\frac{P_a}{2C}(1-\sin\omega_m t),
\end{equation}
where $C$ is the heat capacity of the sample.
This equation has the following analytical solution:
\begin{multline}
\Delta T=\tau_{th}p \left\{1-\frac{\sin (\omega_m t -\phi)}{\sqrt{\omega_m^2\tau_{th}^2 +1}}\right.\\
\left. -\left( 1+\frac{\omega_m \tau_{th}}{\omega_m^2\tau_{th}^2 +1} \right)e^{-\frac{t}{\tau_{th}}}\right\} ,
\end{multline}
where $p=P_{a}/2C$ and $\phi=\arctan(\omega_{m}\tau_{th})$. 
In the $t\rightarrow \infty$ limit, $\Delta T$ oscillates around $\tau_{th} p$ (Fig~\ref{model}(b)).
Since in most cases $M_z$ depends on the temperatures, the magnetization is modulated by the amplitude
\begin{equation}
\Delta M_z^{th}=|\frac{dM}{dT}|\frac{\tau_{th} p}{\sqrt{\omega_m^2\tau_{th}^2 +1}}.
\end{equation}

This model indicates that $\omega_m\tau_{th}$ needs to be smaller than unity to obtain a high signal-to-noise ratio (S/N).
Under real experimental conditions, the measurement requires a long time if $\tau_{th}$ is longer than 1~s.
Moreover, low-frequency measurements are susceptible to vibrational noises.
Therefore, $\tau_{th}\ll 100\ \mathrm{ms}$ is desirable, which can be obtained by maintaining a firm physical contact between the sample and the thermal bath and by using an exchange gas.

Although the principle of the measurement is simple, it requires very different experimental conditions from those of the conventional FDESR. 
In $\Delta M_z^{\mathrm{Bloch}}$ detection, $f_m$ is tuned at the eigenfrequency of the cantilever to improve the force signal, while the measurement is performed in vacuum to maintain high $Q$.
Such conditions correspond to the limit of $\omega_m^2\tau_{th}^2\gg 1$; thus, the contribution of $\Delta M_{th}$ is negligibly small.

\section{Experimental apparatus}
\begin{figure}[tb]
	\begin{center}
		\includegraphics[width=1\hsize]{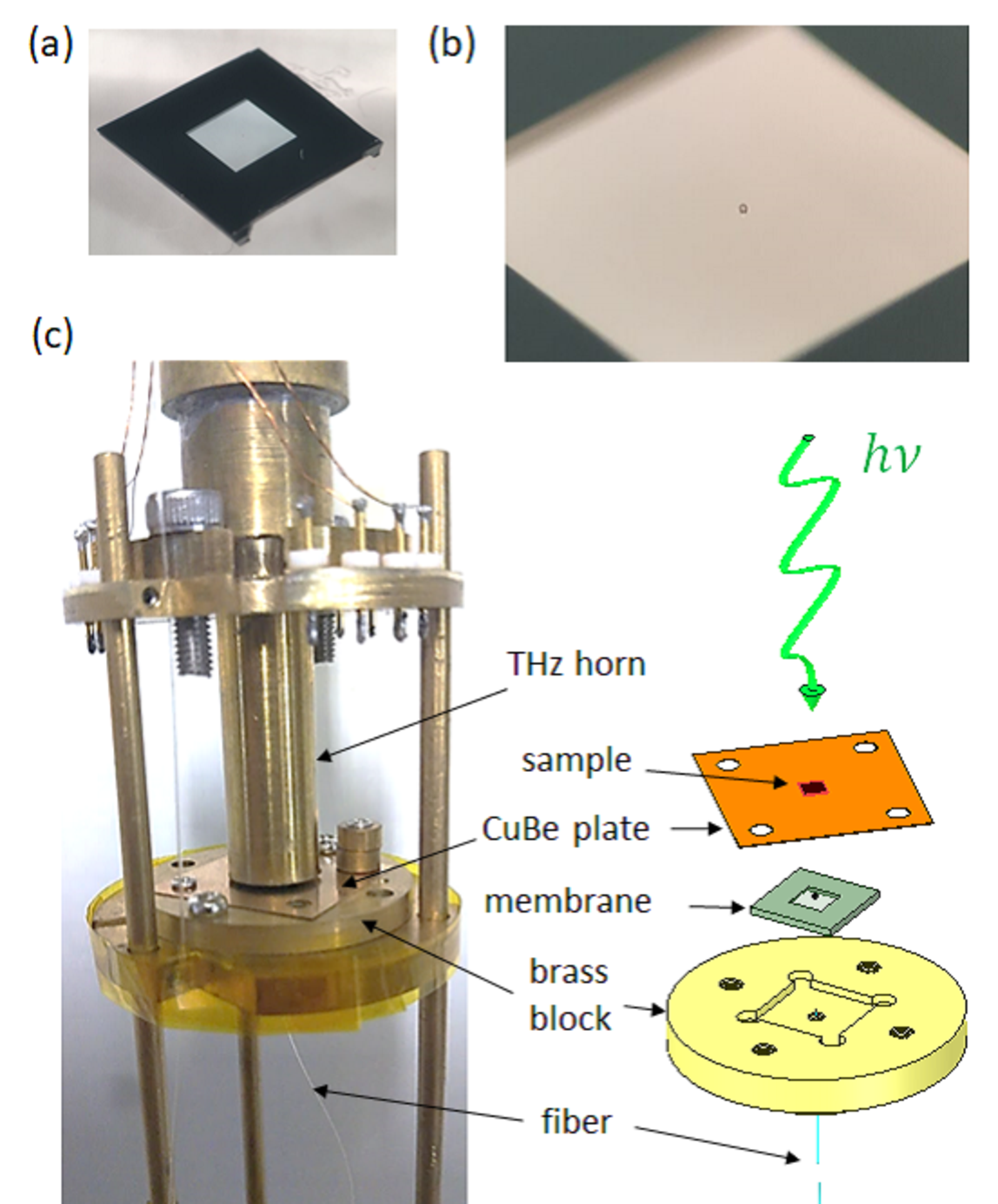}
	\end{center}
\caption{(a) Photo of the SiN$_x$ membrane, (b) SrFe$_{12}$O$_{19}$ particle mounted on the  membrane, and (c) photo and schematic of the experimental setup surrounding the membrane. }
\label{membrane}
\end{figure}

Recently, membrane-type micromechanical devices have proved to be effective for magnetic and~\cite{HT2017,Blankenhorn2017} magnetic resonance measurements~\cite{Scozzaro2016,HT2018}. 
As these devices have excellent physical properties, it is expected that they can address the problems associated with microcantilevers~\cite{Zwickl2008}. 
Figures~\ref{membrane}(a) and \ref{membrane}(b) show the membrane device used in this study (MEM-N03001/7.5M, NTT Advanced Technologies Corporation~\cite{NTTAT}). 
An SiN$_x$ membrane with a thickness $t=$ 100~nm was fabricated on a 3~mm~$\times$~3~mm window on a silicon substrate. 
The eigenfrequency of the lowest mode is $f_0=42.5\ \mathrm{kHz}$. With this value, the internal stress and the spring constant are respectively calculated to be $\sigma=2\rho L^2 f_0^2=$103~MPa and $k=\pi\sigma^2 t/2=$~50.8 N/m, where $\rho=3.1\ \mathrm{g/cm^2}$ is the density of the SiN$_x$. It has been reported that SiN$_x$ nanomembrane has a very high resonance frequency and $Q$-factor. However, $Q$ is in the order of 10--$10^2$ in our setup due to the viscous damping by the heat exchange gas.

We performed experiments by using a ``magnet-on-membrane'' (MOM) configuration in contrast to a previous study using a ``sample-on-membrane'' configuration in the microwave region~\cite{Scozzaro2016}.
A schematic illustration and a photo of the instruments are shown in Fig~\ref{membrane}(c).
We used strontium hexaferrite (SrFe$_{12}$O$_{19}$) whose saturation field is about 2 T as a probe magnet. A spherical particle with a diameter of $\sim$40~$\mathrm{\mu m}$ was glued to the center of the membrane.
Only a small amount of adhesive was used to avoid the degradation of mechanical properties.
We dropped a tiny droplet of Stycast 1266 epoxy using a microinjection pipette with an inner diameter of 20~$\mathrm{\mu m}$.
Typical eigenfrequency and $Q$ of the magnet-mounted membrane are $f_{\mathrm{load}}\sim $1--3~kHz and $Q_{\mathrm{load}}<10$, respectively.

The magnet-mounted chip was covered by a $50\ \mathrm{\mu m}$-thick CuBe plate.
This plate has the following functions in this method:(1) operating as sample stage enabling convenient and fast sample exchange (2) maintaining a firm thermal contact with the brass block to achieve a suitable $\tau_{th}$ value (3) acting as an electromagnetic wave shield to reduce the spurious excitation of the membrane and to prevent the ferromagnetic resonance of the ferrite particle.

The distance between the magnetic particle and the sample was set to $\sim$100~$\mathrm{\mu m}$. 
A magnetic gradient force $\bm{F}_{\mathrm{grad}}=V\nabla(\bm{M}\cdot \bm{B})$, where $V$ is the sample volume, was exerted to the sample by a gradient field generated by a micromagnet.
Consequently, its backaction displaces the membrane.
\begin{figure}[tb]
	\begin{center}
		\includegraphics[width=0.95\hsize]{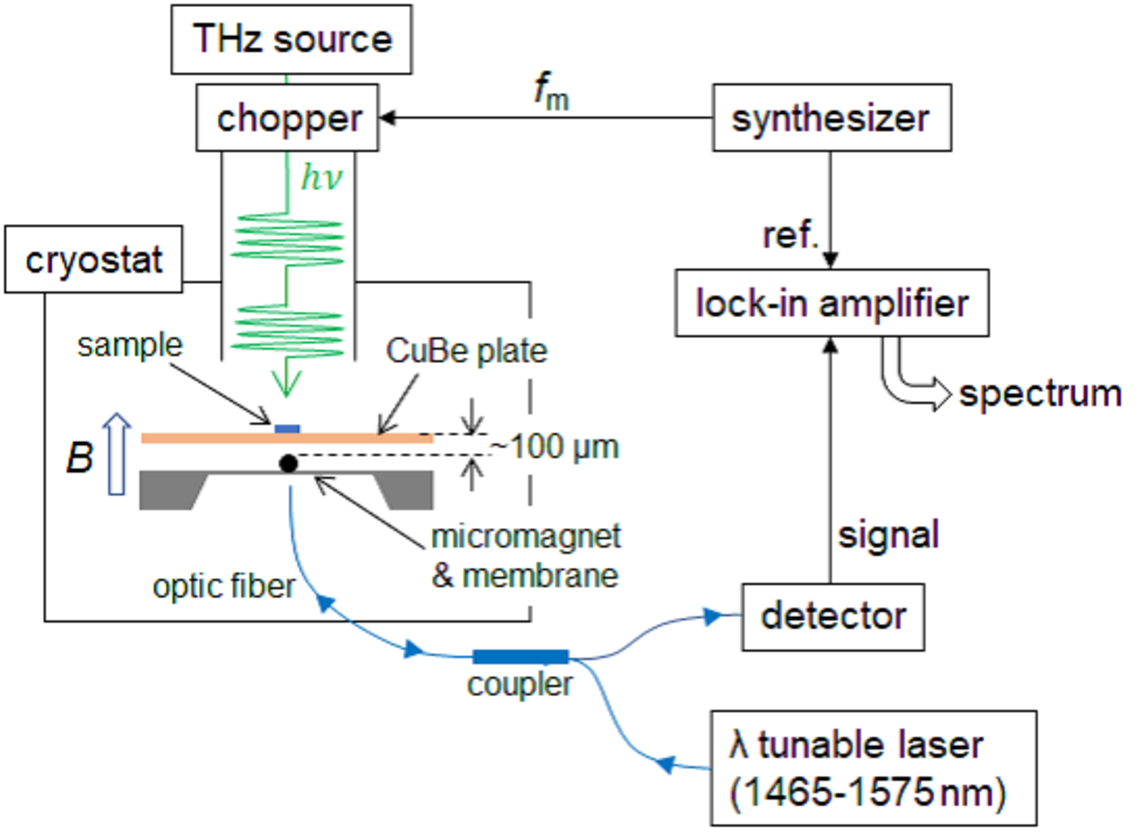}
	\end{center}
\caption{Measurement system with an optical interferometric displacement detection.}
\label{setup}
\end{figure}

The membrane displacement was measured by forming a low-finesse Fabry--Perot interferometer between the membrane surface and the cleaved end of the optic fiber. 
The details of the fiber-optic system are shown in Fig~\ref{setup}.
A fiber-coupled laser (81989A, Agilent) was used as laser source~\cite{Smith2009}.
The reflected beam from the interferometer was divided by the photocoupler (10201A-90, Thorlabs), and fed into a photodetector (81636B, Agilent).
The interference signal, $I$, can be expressed as a function of the cavity length $d$ and the laser wavelength $\lambda$,
\begin{equation}\label{eq1}
I=\frac{I_{MAX}+I_{MIN}}{2}-\frac{I_{MAX}-I_{MIN}}{2}\cos (\frac{4\pi d}{\lambda}),
\end{equation}
where $I_{MAX}$ ($I_{MIN}$) is the maximum (minimum) interference intensity. 
As the wavelength of the 81989A laser source can by varied in the range of 1465--1575~nm, the interferometer can be tuned to the optimal point where $d/\lambda =1/8+n/4\ (n=0,\ 1,\ 2\cdots)$ without using a piezoelectric positioner.

As light sources, we used Gunn diode oscillators (GDOs) below 260~GHz and backward traveling-wave oscillators (BWOs) above 290~GHz. 
The nominal output power depends on the sources, but approximately $P>10\ \mathrm{mW}$ below 160~GHz and $1<P<10\ \mathrm{mW}$ above that frequency.
Their outputs were amplitude modulated, and then propagated through an oversized circular waveguide to the sample space inside the cryostat.
At the bottom of the waveguide, a conical horn was used to focus the electromagnetic wave onto a sample, with an outlet area of $A_{h}=$4~mm$^2$, corresponding to the cutoff frequency of 44~GHz.
$P_a$ is much smaller than $P$, which is calculated taking into account of the attenuation inside the waveguide and horn $\alpha$, the cross-sectional area of the sample perpendicular to $z$-axis $A$, and the absorption efficiency $\epsilon$ determined by the ESR transition probability and thickness. Typical orders of $\alpha$ and $\epsilon$ are $\alpha=-20$--$-10\mathrm{dB}$ and $\epsilon=10^{-3}$--$10^{-2}$. Those values give $P_a=(\alpha \epsilon A/A_h) P=10^{-10}$--$10^{-8}\ \mathrm{W}$ when $A/A_h =10^{-2}$ and $P=1\ \mathrm{mW}$.

The modulation frequency $f_m$ was set to be lower than $1\ \mathrm{kHz}$.
The low $f_{\mathrm{load}}$ and $Q_{\mathrm{load}}$ values are not exhibiting a problem in our method as the ESR signal increases as $f_{m}$ decreases.
When the resonance condition is satisfied, the membrane exhibits a drum-like oscillation, and its signal is detected by a lock-in amplifier with a time constant of 1~s.

\section{Results}
\subsection{DC displacement of the magnet mounted membrane}
Prior to the ESR measurements, DC behavior of the magnet-mounted membrane was analyzed. 
Although the experiment was started after the interferometer was set to the optimal tuning, the DC displacement $\Delta d$ resulting in the change in the sensitivity of the interferometer needed to be considered.
There are two contributing factors to the membrane displacement: the gradient force from the sample and the magnetic torque due to the magnetic anisotropy of the ferrite particle.
We found that in most cases, the latter contribution is dominant.
The magnitude of the magnetic torque changes depending on the position of the ferrite particle~\cite{Blankenhorn2017}. 
Ideally, the displacement at the central position is zero if the magnet is placed in the central position. 
However, as the procedure of mounting the magnet is mainly manual, it is difficult to control the position of the magnet accurately.

Eq. (5) shows that it is favorable that  $\Delta d\ll\lambda/4$ holds during a measurement. In this study, we chose membrane chips which exhibit relatively small displacement under magnetic field. 
If $\Delta d$ deviates from this condition, it is necessary to reset or feedback-control the tuning parameter $d/\lambda$~\cite{HT2016}.

\subsection{Performance test using DPPH measurement}
\begin{figure*}[htb]
	\begin{center}
		\includegraphics[width=1\hsize]{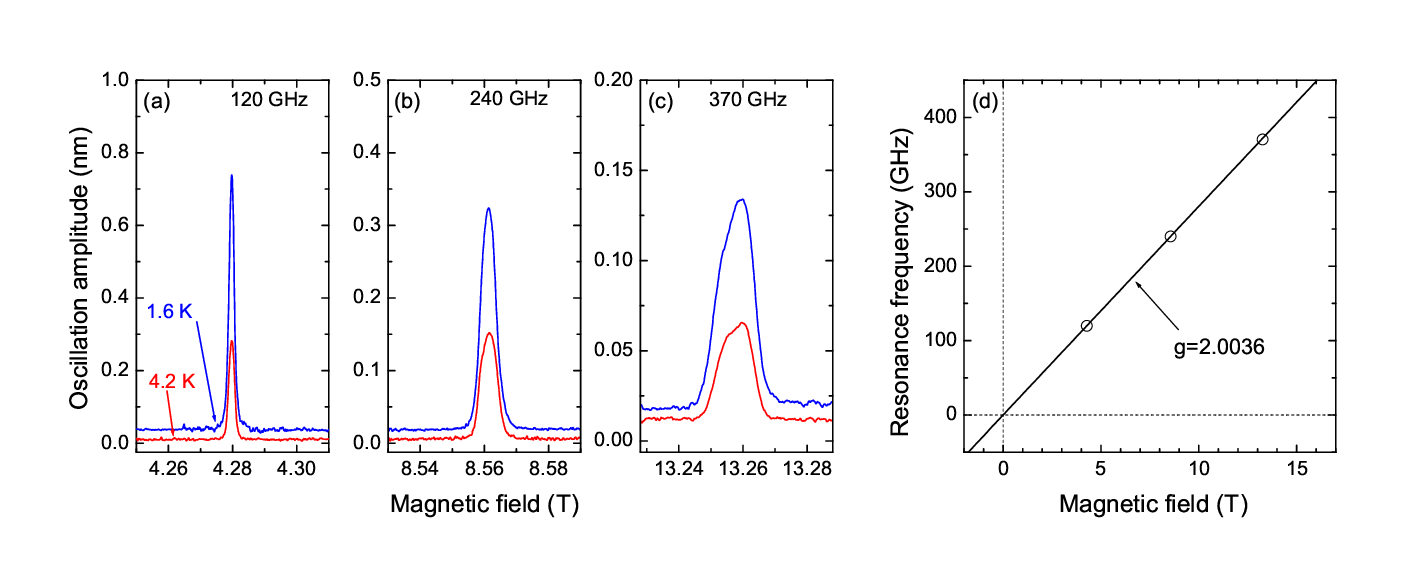}
	\end{center}
\caption{(a)--(c) ESR spectra of DPPH at 120, 240, and 370 GHz, respectively. The magnetic field was swept with a speed of 0.01 T/min.. The data point density is 20 points/mT. (d) frequency vs. magnetic field diagram of DPPH. The black solid line represents a paramagnetic relation with $g=2.0036$.}
\label{DPPH}
\end{figure*}

We measured DPPH for the performance test of the new spectrometer. DPPH is an organic free radical compound and widely used as a magnetic field marker for HFESR measurements, and it is also regularly used as a test sample for FDESR. 
A 20-$\mathrm{\mu g}$ powder sample was mixed with vacuum grease, and then spread on a CuBe plate on an area of approximately $2\ \mathrm{mm}\times 2\ \mathrm{mm}$. 
Figures~\ref{DPPH}(a)-(c) and (d) show the ESR spectra taken at $f_m=13\ \mathrm{Hz}$ and the frequency versus magnetic field diagram.
We observed a sharp absorption peak in the measured frequencies of up to 370~GHz.
It should be noted that we successfully observed ESR signal above 10~T, where the optical detection is technically difficult when a microcantilever is used.
The direction of the membrane displacement is always parallel with the probe beam; therefore, the optical detection can be stably performed even under strong magnetic fields.

In order to use this method for spectroscopy, it is necessary to discuss the possibility of spectrum analysis in addition to peak observation.
In general, force-detected spectrum is distorted from the intrinsic lineshape, because the internal magnetic field inside the sample becomes inhomogeneous when a gradient magnet is used~\cite{Cruickshank2007,Ohmichi2016}.
In our setup, the ferrite particle with the saturation magnetic moment of $m_z=9.1\times10^{-9}\ \mathrm{Am^2}$ produces the local magnetic field $B_z^{\mathrm{local}}=\mu_0 m_z/4\pi z^3$.
At the CuBe plate surface ($z=120\ \mathrm{\mu m}$), $B_z^{\mathrm{local}}=1.05\ \mathrm{mT}$ and $dB_z/dz=26.2\ \mathrm{T/m}$.

We obtained the linewidth of 2.3, 6.0 and 13~mT at 120, 240 and 370~GHz, respectively. These values are larger than the result reported by Krzystek et al.~\cite{Krzystek1997}, $<1$~mT below 500~GHz. However, at the same time, they raised the problem that the linewidth changes depending on the kind of solvent used in the process of crystallization~\cite{Krzystek1997}. 
In fact, other groups reported broader values, 2.5-3.0~mT (93 GHz)~\cite{Tatsukawa1995} and 4.8~mT (250 GHz)~\cite{Lynch1988}.
We consider that the our data include the influence of the uncertain solvent and anisotropic $g$-factor~\cite{Yodzis1963} in addition to the $B_z^{\mathrm{local}}\sim 1$~mT.
In any case, this degree of the broadening by $B_\mathrm{local}$ does not affect the analysis of the broad spectrum which we are intersted in.

\begin{table*}[tb]
\caption{S/N and spin sensitivity for the spectra shown in Figs.~\ref{DPPH}(a)--(c). $B_1$ and $|\Delta M_z^{\mathrm{Bloch}}/M_0|$ values were calculated assuming 10~dB attenuation inside the waveguide for simplicity.}
\centering
\begin{tabular}{cccccc}
frequency (GHz)  & \hspace{5pt}S/N & \hspace{5pt}sensitivity($\mathrm{spins/G\cdot s}$)& \hspace{5pt}power (mW)& \hspace{5pt} $B_1$ (T)&\hspace{5pt}$|\Delta M_z^{\mathrm{Bloch}}/M_0|$\\
& \hspace{5pt} $T=4.2\ \mathrm{K}$, $T=1.6\ \mathrm{K}$& \hspace{5pt}$T=4.2\ \mathrm{K}$, $T=1.6\ \mathrm{K}$ & \hspace{5pt}& \hspace{5pt}\\
\hline
\hline
\hspace{10pt}120 & \hspace{10pt}210, 270 & \hspace{10pt}$6.4\times 10^{12}$, $5.0\times 10^{12}$ & 36 & $1.1\times 10^{-6}$ & $5\times 10^{-4}$\\
\hspace{10pt}240  & \hspace{10pt}140, 430 & \hspace{10pt}$3.5\times 10^{12}$, $1.1\times 10^{12}$ & 2 &  $2.6\times 10^{-7}$ & $3\times 10^{-5}$\\
\hspace{10pt}370  & \hspace{10pt}91, 120 & \hspace{10pt}$2.4\times 10^{12}$, $1.9\times 10^{12}$ & 10 &  $5.8\times 10^{-7}$ & $1\times 10^{-4}$\\
\end{tabular}
\label{table}
\end{table*}

The obtained S/N and spin sensitivity are summarized in Table~\ref{table}. 
It should be noted that the spin sensitivities shown here were calculated by using the total sample mass.
As the decrease in the field gradient follows $r^{-4}$ as the distance from the ferrite particle $r$ increases, the force signal originates in a limited region close to the CuBe plate.
Therefore, this calculation gives only the lower bounds of the spin sensitivity, whereas the highest sensitivity of $1.1\times 10^{12}$~spins/G$\cdot$s was obtained at 240~GHz and $T=1.6\ \mathrm{K}$.
Although the nominal output power is the highest at 120~GHz, the spin sensitivity at this frequency is the lowest among the three measured frequencies.
This is probably due to the frequency-dependent return loss at the conical horn.
We estimated that the highest electromagnetic wave power at the horn outlet is in the order of 1~mW at these frequencies.

\subsection{Background signal due to spurious oscillation}
It can be seen that the finite membrane oscillations were observed even in the off-resonant field region in Figs.~\ref{DPPH}(a)--(c).
The offset signal originates from the spurious mechanical oscillation, which often causes problems in FDESR.
It is also referred to as ``modulation noise'' as it occurs when the micromechanical device couples with the modulation source.
A few techniques have been proposed to suppress the spurious excitation, such as harmonic detection~\cite{Rugar1992} and anharmonic modulation~\cite{Bruland1995}.
However, these techniques are not effective in our case owing to the low quality factor ($Q\simeq 1$) of the magnet-mounted membrane and the shift of the resonant frequency in the magnetic field.

In the THz region, the main source of the spurious noise is the amplitude-modulation of the electromagnetic wave, which provides thermal force to the micromechanical device and strongly excites its oscillation. 
Its effect was apparent in our previous experiment using a sample-on-cantilever (SOC) configuration. Spurious excitation does not only increase the background signal, but also, in more severe cases, results in a reversal of polarity of the signal~\cite{HT2015,HT2016}.
In the current design, the shielding structure prevents the membrane to be irradiated directly by the electromagnetic wave.
In addition, by setting $f_m$ far below $f_{\mathrm{load}}$, the spurious excitation can be reduced. Although the spurious level is still much higher than other noise sources, such as thermal oscillation noise ($=\sqrt{k_\mathrm{B} T/k}\sim 1$~pm at $f_\mathrm{load}$) expected from the equipartition theorem, it becomes nearly independent of the magnetic field. In Figs.~\ref{DPPH}(a)--(c) with lock-in time constant 1s, the baseline noise has an peak-to-peak amplitude of about 3 pm and the standard deviation of about 1pm. Empirically, these values vary by a factor of few depending on detailed experimental conditions (e.g. magnet particle size, fiber alignment.).

\begin{figure}[tb]
	\begin{center}
		\includegraphics[width=0.95\hsize]{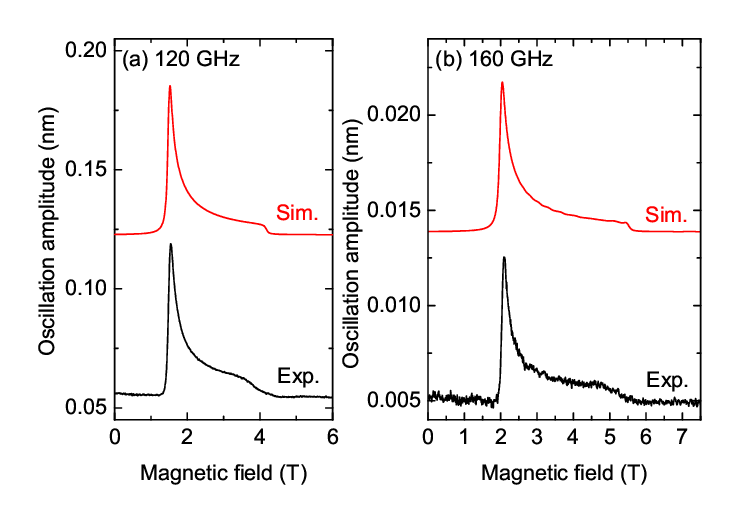}
	\end{center}
\caption{ESR spectra of hemin chloride powder at (a) 120 and (b) 160~GHz. The black and red curves show the experimental and simulated spectra, respectively. The magnetic field was swept with a speed of 0.5~T/min. The data point density is 400 points/T}. The simulation was performed by using the EasySpin software~\cite{Stoll2006}. We used the following parameters determined by a transmission-type method under pulsed magnetic field~\cite{Okamoto2018}: $g_{x}=g_{y}=1.93$, $g_{z}=2.05$, anisotropic parameters $D=6.90\ \mathrm{cm^{-1}}$ and $E=0.055\ \mathrm{cm^{-1}}$.
\label{hemin}
\end{figure}

Such stable background is particularly important for observing signals with broad linewidth.
Figure~\ref{hemin} shows the ESR spectrum of 0.7~mg of hemin chloride C$_{34}$H$_{32}$ClFeN$_{4}$O$_{4}$ powder, which is a metal porphyrin complex with a square-planar molecular structure~\cite{Okamoto2018}. 
The magnetic ion of hemin is Fe(III) in the high spin state ($S=5/2$). 
As it has anisotropic $g$-values, a broad ESR spectrum was observed for the powder sample between the fields corresponding to $g^{eff}_{\mathrm{para}}\simeq 2$ and $g^{eff}_{\mathrm{perp}}\simeq 6$, where $g^{eff}_{\mathrm{para}}$ and $g^{eff}_{\mathrm{perp}}$ are the effective $g$-values when $\bm{B}$ is applied parallel to and perpendicular with the normal of the planar structure, respectively~\cite{Okamoto2018}.
The reasonable agreement with the simulated curve proves that a very broad ESR signal can be observed without losing the spectral information.
This is in contrast with the ESR spectra taken by the SOC technique, where the spectral information was lost, except for the main peak at $g^{eff}_{\mathrm{perp}}\simeq 6$, due to the large field-dependent background~\cite{HT2016}.

\subsection{Application to quantum spin materials}

\begin{figure*}[tb]
	\begin{center}
		\includegraphics[width=0.95\hsize]{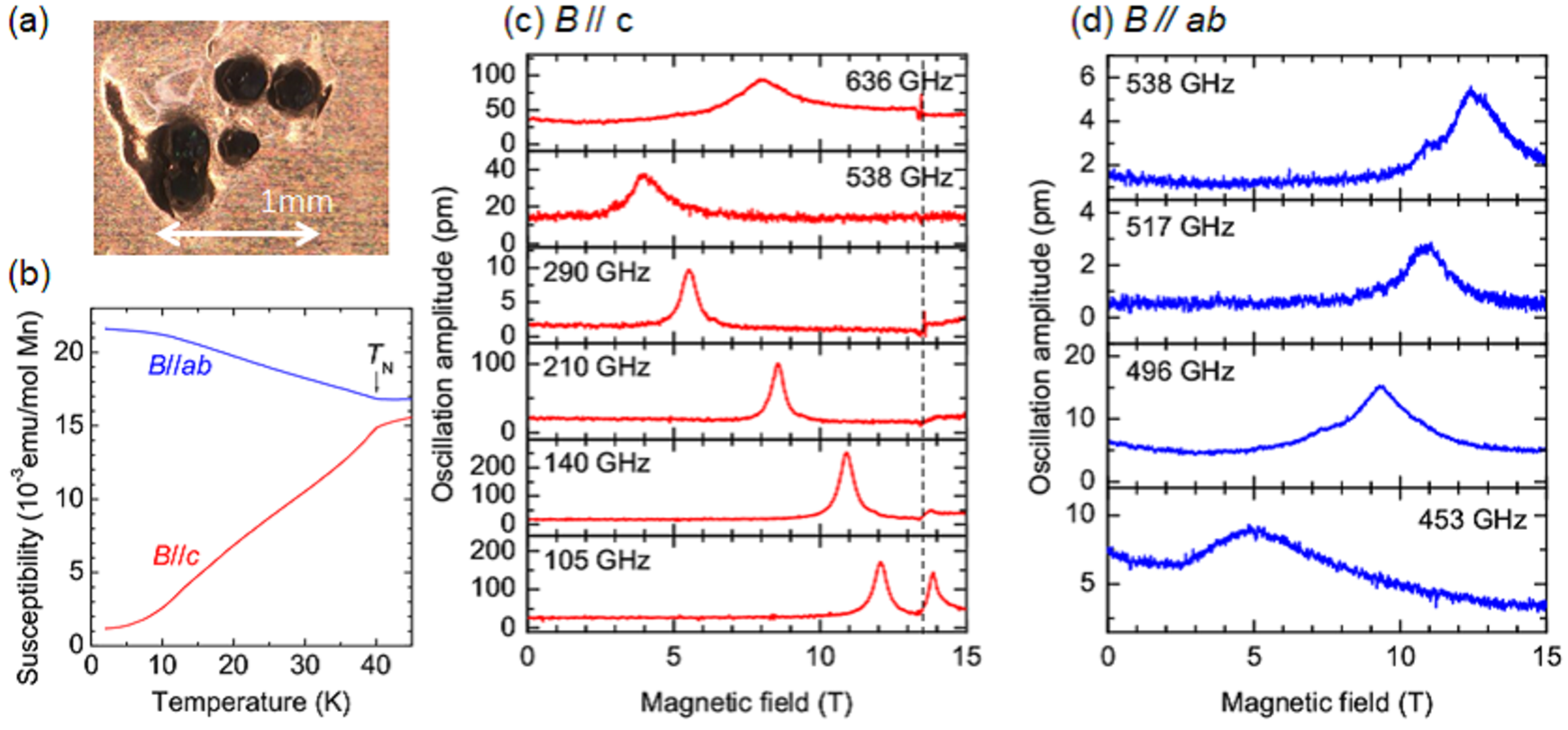}
	\end{center}
\caption{(a) Photo of KMn$_3$Ge$_2$O$_9$ single crystals on the CuBe plate. The $c$-axis is pointing out of the plane of the figure for $\bm{B}\parallel c$ measurements. (b) Temperature dependence of the sample magnetization, (c) and (d) ESR spectra of KMn$_3$Ge$_2$O$_9$ for (c) $\bm{B}\parallel c$ and (d) $\bm{B}\parallel ab$. The magnetic field was swept with a speed of 0.5 T/min.. The data point density is 400 points/T}. The dashed line in (c) indicates the magnetic field (13.5~T) where the compound undergoes a spin-flop transition.
\label{KMGO}
\end{figure*}

\begin{figure}[tb]
	\begin{center}
		\includegraphics[width=0.85\hsize]{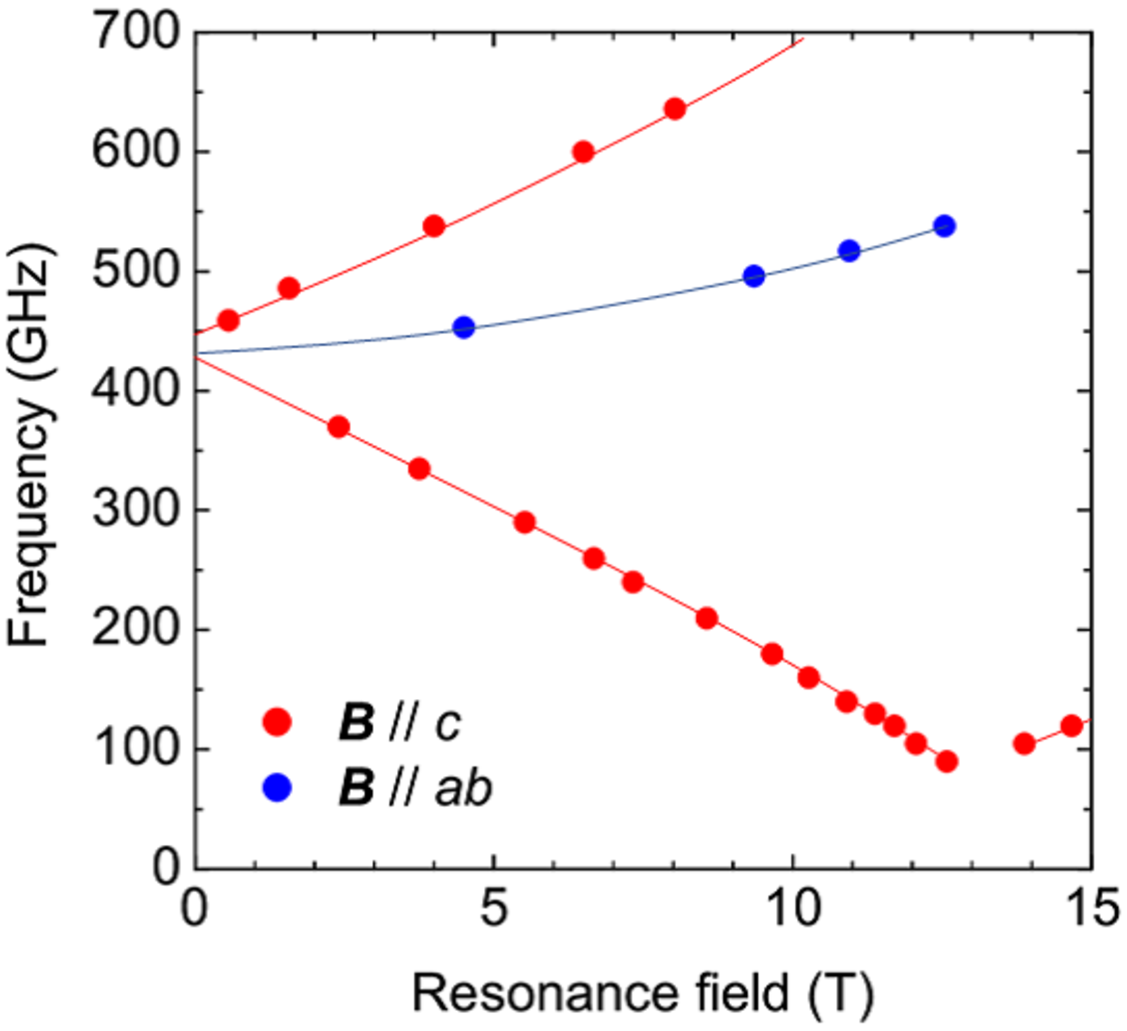}
	\end{center}
\caption{Frequency vs. magnetic field diagram of KMn$_3$Ge$_2$O$_9$.}
\label{KMGO_FFdiagram}
\end{figure}

In certain magnetic materials, interesting phenomena emerge as a result of quantum fluctuation, geometrical frustration, and a topological feature~\cite{Balents2010, Hikihara2008, Haldane1983}, which cannot explained by classical theories.
As the HFESR can directly study the spin dynamics and the magnetic interaction between spins, it is effective in the research of quantum spin systems.
However, the conventional transmission-type method requires a sufficient amount of sample (typically at least 1~mg), which is often a problem in the study of newly found materials.

The improvement in sensitivity is also important in another sense.
It is expected that the quality of the HFESR spectrum is improved by reducing the sample size.
When a large single crystal sample is measured by a transmission method, the shape of the spectrum is often distorted due to the sudden change in the refractive index accompanying the resonance and the interference of the electromagnetic waves inside the sample. 
In order to avoid these problems, it is necessary to cut the sample into a wedge shape or to shape the sample sufficiently smaller than the wavelength. 
Obviously, these procedures reduce the S/N.

In the previous section, we demonstrated a successful ESR detection method above 10~T. 
This high sensitivity of the novel spectrometer can be applied to the study of magnetic materials.
For the next example, we present an ESR study of recently discovered Kagome lattice antiferromagnet, KMn$_3$Ge$_2$O$_9$, synthesized by hydrothermal method~\cite{Hara2018}.
This compound has a hexagonal unit cell with a $P6_3/mmc$ space group, and lattice constants of $a=5.833\ \mathrm{\AA}$ and $c=13.711\ \mathrm{\AA}$.
The Mn$^{3+}$ ($S=2$) ions located at the center of the MnO$_6$ octahedra form a Kagome lattice layer along the $c$-plane.
It has been confirmed by magnetization and heat capacity measurements that the Mn$^{3+}$ has an antiferromagnetic order at temperatures below 40~K, although the details of the magnetic structure are still unclear. 
Five pieces of small single crystals with a total mass of 0.2~mg $A/A_h=1.4\times 10^{-2}$ were placed on the CuBe plate as shown in Fig.~\ref{KMGO}(a).
Then an ESR measurement was performed at 4.2~K in $\bm{B}\parallel c$ and $\bm{B}\parallel ab$ configurations.

Figures~\ref{KMGO}(c) and \ref{KMGO}(d) show the ESR spectra of the KMn$_3$Ge$_2$O$_9$ for $\bm{B}\parallel c$ and $\bm{B}\parallel ab$ measurements. 
Despite the low mass sample, the ESR signal was clearly observed at each frequency.
The anomaly around 13.5~T is due to the spin-flop transition where we found a jump in the DC displacement of the membrane.
The highest $\mathrm{S/N}=2.0\times 10^2$ was obtained at 140~GHz, where the spin sensitivity was calculated as $1.4\times 10^{12}$ spins/G$\cdot$s.

The results are summarized in Fig.~\ref{KMGO_FFdiagram}.
The significant advantage of HFESR is that it can determine the magnetic anisotropy of an antiferromagnet definitively.
The frequency-dependence of the resonance field shows a typical behavior of an easy axis-type antiferromagnet with a spin-easy axis in the direction of the $c$-axis.
This result confirms the expectations from the results of the magnetization measurement~\cite{Hara2018}.

\section{Discussion}
\subsection{Evidence of thermally modulated magnetization}
Using the novel membrane-based spectrometer, we obtained a moderately high spin sensitivity regardless of $T_{1(2)}$.
In this section, we provide a quantitative argument to clarify the dominant observations of $\Delta M_z^{th}$ in the abovementioned experiments.

The change in the $z$ component of the gradient force due to ESR can be  expressed as
\begin{equation}
\Delta F_{\mathrm{grad}}=V_{\mathrm{slice}}\Delta M_z\frac{\partial B_z}{\partial z},
\end{equation}
where $V_\mathrm{slice}$ is the volume of the resonance slice. 
In magnetic resonance force microscopy applications, $V_{\mathrm{slice}}$ is limited by the large field gradient to a very narrow region within the sample.
In contrast, we used $dB/dz=\leq 26.2\ \mathrm{T/m}$, which is significantly smaller.
This suggests that the slice thickness, $t=\delta B/(\frac{\partial B_z}{\partial z})$ is so large that it covers the entire sample.
In this case, $\Delta F_{\mathrm{grad}}/F_{\mathrm{grad}}$ is equal to $\Delta M_z /M_0$.
In the DPPH (KMn$_3$Ge$_2$O$_9$) measurements, we observed an oscillation amplitude of 0.30~nm at 120~GHz (0.25~nm at 140 GHz) and $T=4.2 \ \mathrm{K}$, corresponding to $\Delta F_{\mathrm{grad}}= 15\ \mathrm{nN}\ (13\ \mathrm{nN})$.
Considering $F_\mathrm{grad}$, although we could not extract its contribution from the DC displacement, it was calculated to be smaller than $4.9 \ \mathrm{\mu N}$ ($3.0 \ \mathrm{\mu N}$) using its magnetic susceptibility of $8.6\times 10^{-2}\ \mathrm{emu/mol}$ ($1.3\times 10^{-3}\ \mathrm{emu/mol(Mn)}$ when $B\parallel c$).
Therefore, $|\Delta M_z /M_0 |>3.1\times 10^{-3}$ ($4.3\times 10^{-2}$).
This value is too large to originate from $\Delta M_z^{\mathrm{Bloch}}$.
As summarized in Table. I, our low power sources can induce only small $|\Delta M_z^{\mathrm{Bloch}}/M_0 |\sim 10^{-4}$--$10^{-5}$ for DPPH.
This discrepancy is more pronounced in KMn$_3$Ge$_2$O$_9$ where $T_{1(2)}$ is 3--4 orders shorter than that of DPPH.

The change in thermal magnetization is a more plausible source of such a large magnetization modulation.
Both DPPH and KMn$_3$Ge$_2$O$_9$ have temperature-dependent magnetization. At $T=4.2\ \mathrm{K}$, $dM/dT=2.0\times 10^{-2} \ \mathrm{emu/mol\cdot K}$ ($1.0\times 10^{-4}\ \mathrm{emu/mol(Mn)\cdot K}$) for DPPH (KMn$_3$Ge$_2$O$_9$ when $B\parallel c$).
Therefore, if the sample is heated up by 40 mK (130 mK), it would induce 1~\% change of the total magnetization.
As for KMn$_3$Ge$_2$O$_9$, we can calculate $P_a$ corresponding to $\Delta T=130$~mK to be $P_a=4.7\times 10^{-10}$~W using its specific heat $1.35\times 10^{-6}\ \mathrm{J/K\cdot mol(Mn)}$. This result shows a good agreement with the value estimated in our setup (Sec. III).

\begin{figure*}[tb]
	\begin{center}
		\includegraphics[width=0.75\hsize]{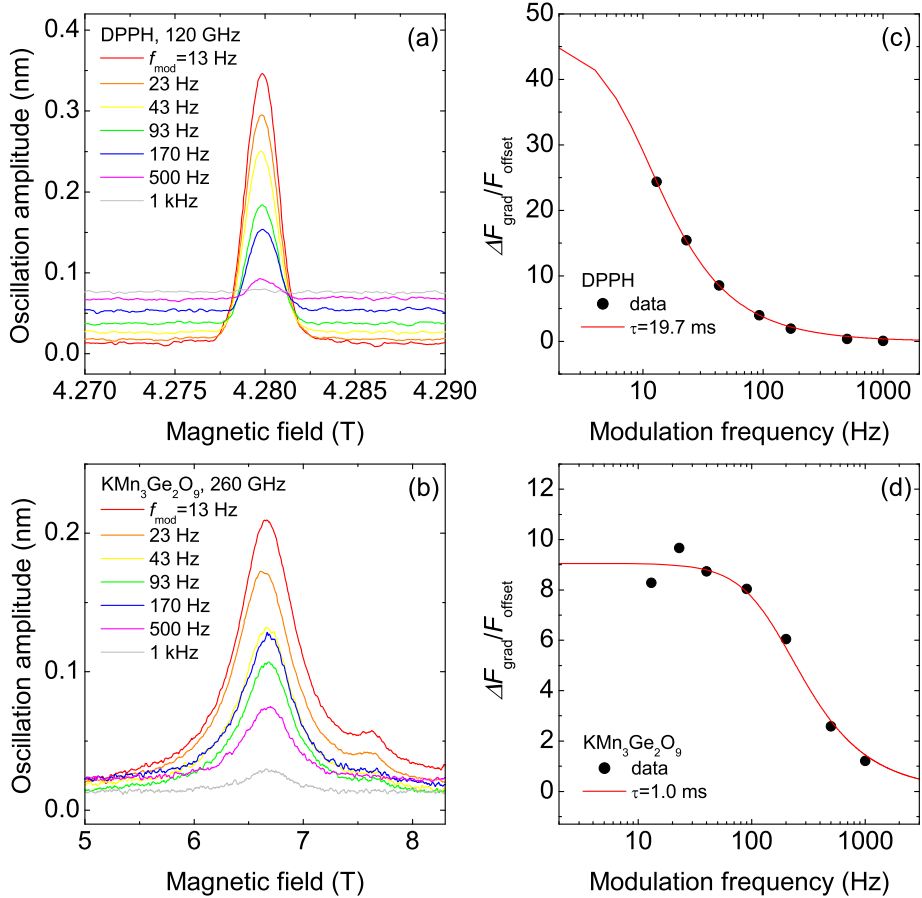}
	\end{center}
\caption{(a) and (b) ESR spectra of the DPPH at 120~GHz and the KMn$_3$Ge$_2$O$_9$ at 260~GHz taken with different $f_m$s. The minor structure around 7.5~T in (b) is probably due to the impurity phase. (c) and (d) Modulation frequency dependence of the force signal normalized by the offset signal. The black closed circles and the red solid lines show the experimental data and the best fit for Eq. (4), respectively.}
\label{mod_dependence}
\end{figure*}
A more obvious evidence for the thermally modulated magnetization is the $f_m$-dependence of the ESR signal (Figs.~\ref{mod_dependence}(a) and (b)).
In general, the larger oscillation amplitudes were observed as $f_m$ decreases. We divided the spectra into the two components, ESR-induced oscillation and the spurious excitation that produces field-independent offset signal. By taking the ratio of two components, we can discuss the frequency dependence of $\Delta F_{\mathrm{grad}}$ with excluding the frequency response of the membrane (Figures~\ref{mod_dependence}(c) and (d)).
The data can be well fitted by the function $\propto 1/\sqrt{\omega_m^2 \tau_{th}^2 +1}$ suggested by Eq. (4). 
The ESR signal of the DPPH was totally suppressed at $f_m =1\ \mathrm{kHz}$, which indicates that $M_z^{\mathrm{Bloch}}$ is below the detection limit.
Values $\tau_{\mathrm{th}}=19.7$ and $1.0\ \mathrm{ms}$ were derived for the data for the DPPH and the KMn$_3$Ge$_2$O$_9$, respectively.
The difference in $\tau_{\mathrm{th}}$ values is considered to be the result of the difference in the qualities of the thermal contact between the sample and the CuBe plate.

This technique has not been applied to temperatures above 4.2~K. 
However, our model shows that the signal intensity decreases with the increasing temperature, because $\Delta T$ is inversely proportional to the heat capacity.
For a paramagnetic sample, the decrease in $|dM/dT|$ ($\propto T^{-2}$) also results in the degradation of sensitivity.
Conversely, this approach can be effective in measurements at lower temperatures.
The $dM/dT$ value of KMn$_3$Ge$_2$O$_9$ is not particularly large among antiferromagnets; therefore, this method is considered to be applicable to several magnetic materials.

\subsection{Comparison with other techniques}
Force-detected HFESR has been studied in various configurations (e.g. SOC~\cite{Ohmichi2008,Hallak2010,HT2015,Ohmichi2016,HT2016}, magnet-on-cantilever~\cite{Cruickshank2007}, and sample-on-membrane~\cite{HT2018}).
Among these, the SOC configuration achieved the highest absolute spin sensitivity in the range of 10$^8$--10$^9$spins/G$\cdot$s.
However, as it was indicated in Section 1, the SOC is only effective under limited conditions due to certain technical issues.
The new method is significant in the sense that it solved these problems and enabled the use of the FDESR method at a practical level.
We believe that the achieved spin sensitivity of $\sim 10^{12}$ spins/G$\cdot$s is sufficient in most cases.
A further advantage is that the MOM configuration enables the increase in the amount of the sample to improve the S/N.
However, it should be noted that the S/N does not increase proportionally with the sample volume.
The magnetic field gradient generated by the ferrite particle is nearly completely attenuated within a distance of $\sim$1~mm.
Therefore, the transmission-type method is more suitable for low spin concentration samples, such as solution samples.

We believe that the most suitable application of this method is in submillimeter-sized single crystal measurements, as presented in Section IV.
In the case of transmission-type measurements, when the sample is much smaller than the cross-sectional area of the waveguide, most of the terahertz-wave is transferred from the sample space without irradiating the sample.
In order to improve the S/N, many pieces of single crystals need to be placed at the sample space with their crystal axes aligned. 
As the number of the crystals increases, the misalignment of the crystal axes becomes large, and the advantages of the single crystal measurement over the polycrystalline sample measurement are lost.
Therefore, the new method is a powerful tool for the fast and detailed characterization of newly discovered materials.

\section{Summary}
In this study, we reported recent developments in the HFESR technique with two of our significant achievements: (i) Development of an SiN$_x$-nanomembrane-based HFESR spectrometer, which overcomes the technical problems of cantilever technique and
(ii) detection of thermally modulated magnetization, which enables the robust measurement independent of the material-dependent spin relaxation time.
By combining these techniques, we demonstrated broadband HFESR spectroscopy of DPPH, hemin chloride, and antiferromagnetic KMn$_3$Ge$_2$O$_9$ in magnetic fields of up to 15~T and frequencies of up to 636~GHz with a spin sensitivity of $10^{12}$~spins/G$\cdot$s. 

Our apparatus is the first practical force-detected HFESR spectrometer. 
By reducing the magnetic field gradient, it is possible to observe a spectrum without large distortion even for materials with a narrow line width. 
Moreover, the low spurious noise stabilizes the background signal, such that it became possible to analyze a spectrum with a wide line width, such as a non-oriented spectrum of anisotropic substances and a broad absorption line of strongly interacting spin systems. 
Our method enables fast characterization of new substances including fast-relaxing magnetic materials, and it is expected to contribute to the discovery of new quantum phenomena.

\section{Acknowledgement}
This study was partly supported by the Grant-in-Aid for Young Scientists (B) (16K17749) and the Grant-in-Aid for Scientific Research (B) (No. 26287081) from the Japan Society for the Promotion of Science and the Asahi Glass Foundation.

\end{document}